\documentclass[11pt,twoside]{article}
\usepackage{asp2010}

\resetcounters

\begin{document}

\title{Distributed GPU Volume Rendering of ASKAP Spectral Data Cubes}
\author{A. H. Hassan$^1$, C. J. Fluke, and D. G. Barnes
\affil{Centre for Astrophysics and Supercomputing, Swinburne University of Technology, PO Box 218, Hawthorn, Australia, 3122. 
\affil{$^1$ahassan@swin.edu.au}}
}

\begin{abstract}
The Australian SKA Pathfinder (ASKAP) will be producing 2.2 terabyte HI spectral-line cubes for each 8 hours of observation by 2013. Global views of spectral data cubes are vital for the detection of instrumentation errors, the identification of data artefacts and noise characteristics, and the discovery of strange phenomena, unexpected relations, or unknown patterns. We have previously presented the first framework that can render ASKAP-sized cubes at interactive frame rates. The framework provides the user with a real-time interactive volume rendering by combining shared and distributed memory architectures, distributed CPUs and graphics processing units (GPUs), using the ray-casting algorithm.
In this paper we present two main extensions of this framework which are: using a multi-panel display system to provide a high resolution rendering output, and the ability to integrate automated data analysis tools into the visualization output and to interact with its output in place. 

\end{abstract}

\section{Intoduction}
Upcoming radio observation facilities such as  the Australian Square Kilometre Array Pathfinder (ASKAP), the MeerKAT Karoo Array Telescope, the Low Frequency Array (LOFAR), and ultimately the Square Kilometre Array (SKA) will pose a significant challenge for current astronomical data analysis and visualization tools. The expected data product sizes (e.g. 2.2 TB per 8 hour of observation for the proposed WALLABY all-sky HI survey\footnote{See http://www.atnf.csiro.au/research/WALLABY for details.}) are orders of magnitude larger than astronomers, and existing astronomy software, are accustomed to dealing with. 

In \citet{Hassan:2010a}, we presented a distributed GPU framework to interactively volume render larger-than-memory astronomical data cubes. Throughout this work, we demonstrated how volume rendering offers an alternative to standard 2D visualization techniques and provided a way to overcome the technological barrier caused by the computational requirement of volume rendering for large data cubes. The presented framework utilizes a heterogeneous CPU and GPU hardware infrastructure, combining shared- and distributed-memory architectures, to yield a scalable volume rendering solution, capable of volume rendering image cubes larger than a single machine memory limit, in real-time and at interactive frame rates. The usage of GPUs as the main processing backbone for the system, and the remote visualization architecture adopted in our design for this framework enables further enhancement for the astronomer's visualization experience.

In this paper we present an extension to this framework to provide: better visualization output by integrating external quantitative information with the volume rendering output (see section \ref{sct:Duchamp}) , and high resolution output via multi-panel displays (see section \ref{sct:Opti}).

\section{Integrating Duchamp Output}
\label{sct:Duchamp}
While the main processing burden is moved to the back-end cluster of GPUs, the rendering client is free to do some relatively smaller processing to enhance the interactivity and the scientific outcomes of the visualization output. The usage of remote visualization enables the viewer application to integrate different graphical primitives with the volume rendering output without affecting the rendering processes  [see Figure 2 in \citet{Hassan:2010a} for details of the framework's main components]. We already utilized this ability to present simple primitives such as the colour map, and orientation arrows in the original implementation [see Figure 1, 5, and 6 in \citet{Hassan:2010a}]. We have extended our work to enable the overlay of a source finder output catalogue on the volume rendering output. We use the Duchamp source finder \citep{whiting:2008}, the selected source finder for the proposed ASKAP survey pipeline, to demonstrate this integration.

\begin{figure*}
\begin{center}
\includegraphics{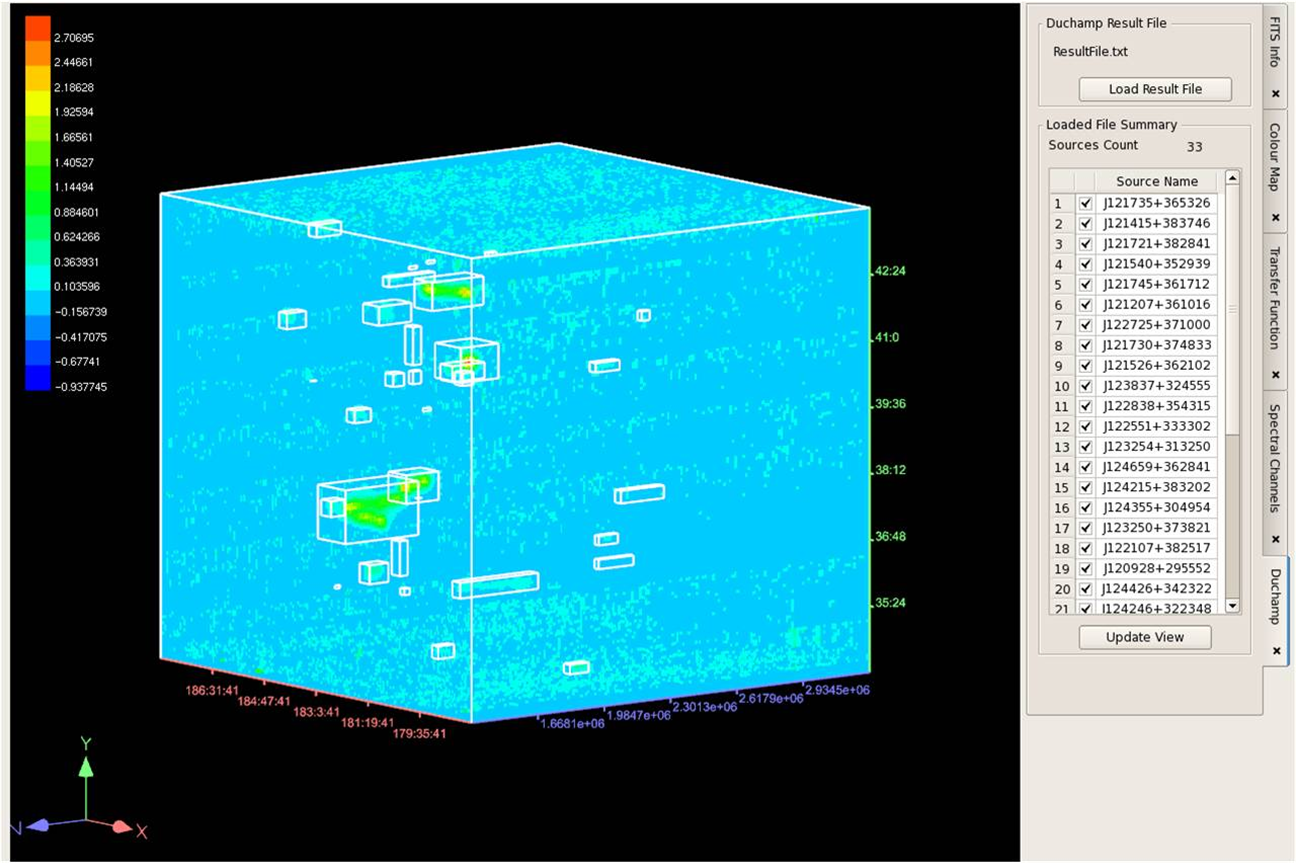}
\caption{Illustration of the process of integerating Duchamp catalogue with the volume rendering output. The data cube used is a neutral hydrogen (21cm) observation of part of the Ursa Major galaxy cluster, made with the Lovell radio-telescope at Jodrell Bank, Manchester. Data courtesy Virginia Kilborn (Swinburne).
}
\label{fig:Duchamp}                                 
\end{center}                                 
\end{figure*}

First, the user loads a Duchamp output file - an ASCII file that contains a set of detected astronomical sources represented by their bounding cubes. The viewer application parses the file and extracts the sources' bounding cubes, which are superimposed as wireframe boxes on the volume rendering output. The user interaction with the displayed volume rendering output is automatically applied to the displayed source cubes. The user can select one of the source cubes to display further available information about that source. See Figure \ref{fig:Duchamp} for an illustration of this process. The user can also show/hide a group of the displayed sources to enable better understanding of the output. We anticipate other applications of this extension such as comparing multiple source finder outputs (by assigning different colour for the source cubes), and overlaying radio/optical catalogues.

\section{OptIPortal Integeration}
\label{sct:Opti}
\begin{figure}
\begin{center}
\includegraphics{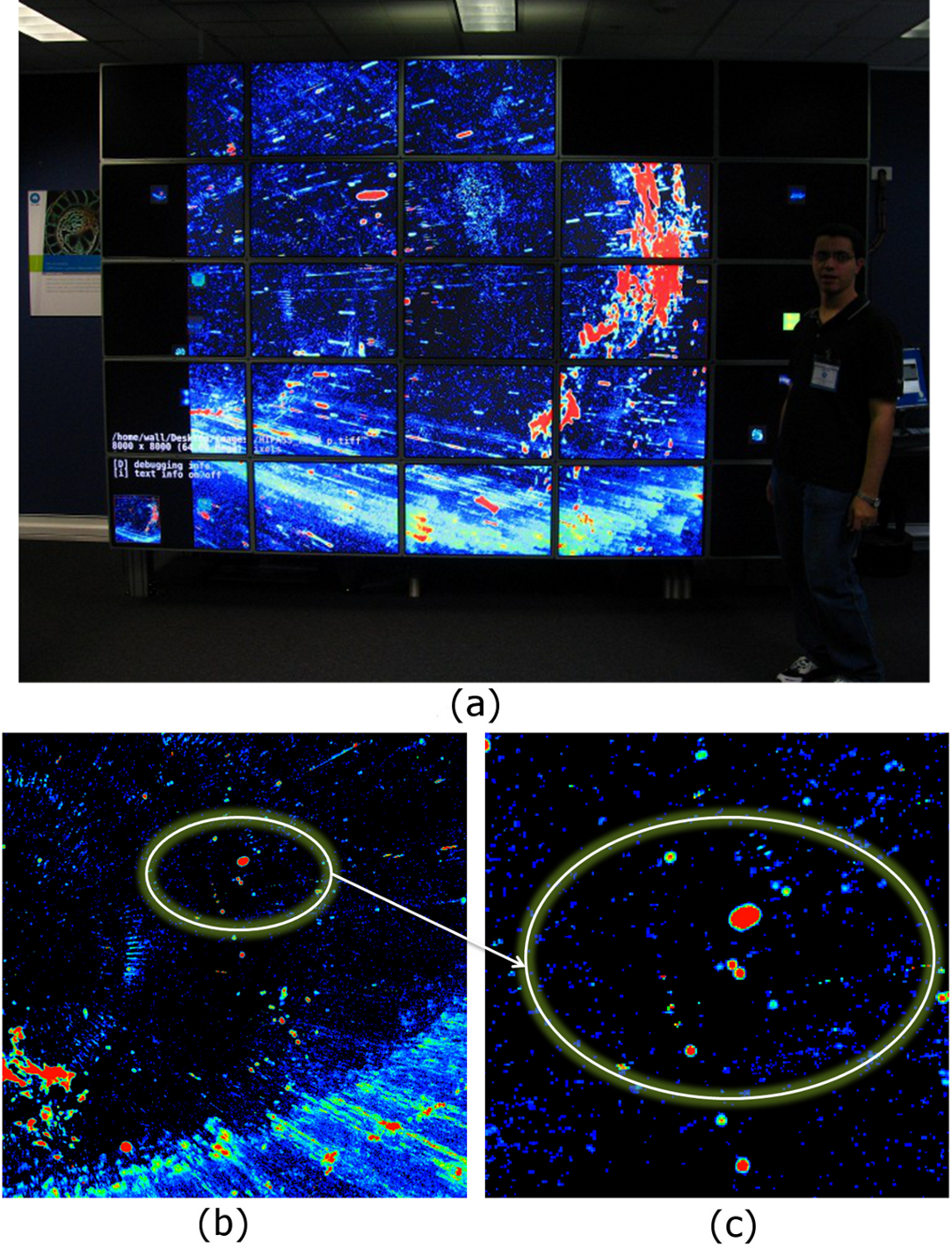}
\caption{(a) The HIPASS cube in 8000 $\times$ 8000 pixel resolution displayed using the CSIRO OptIPortal ATNF, Marsfield (9600 $\times$ 5400 pixel). Credit: C. J. Fluke. (b) A portion of the HIPASS volume rendering with 3000 $\times$ 3000 output resolution. (c) The same portion of the HIPASS cube with 8000 $\times$ 8000 otuput resolution.  }
\label{fig:Resolution}                                 
\end{center}                                 
\end{figure}

Another benefit of using GPUs is the ability to render high resolution output in a relatively short time.  So, as a further extension of the current viewer implementation, we present support for high resolution tiled display systems (such as the OptIPortal display\footnote{\url{http://research.ict.csiro.au/research/labs/information-engineering/ie-lab-projects/optiportal}}). We used   the OptIPortal display facility operated by the Australian Commonwealth Scientific and Research Organization (CSIRO) at Marsfield (Sydney, Australia) to demonstrate this integeration (see Figure \ref{fig:Resolution}-a). The current CSIRO OptIPortal consists of 5 $\times$ 5 high definition screens (1920 $\times$ 1080 pixel) with an overall output size of 9600 $\times$ 5400 pixels. The viewer uses the back-end GPU cluster to render a static image of the current data file with the same orientation information but with a different output size (defined by the user). The rendering output is stored in TIFF format, which is converted to the pyramidal tiled TIFF format\footnote{See \url{http://iipimage.sourceforge.net/documentation/images/}}.  

Figure \ref{fig:Resolution}-b and \ref{fig:Resolution}-c shows an illustration of the importance of high resolution output with the HIPASS southern sky cube\footnote{The southern sky cube was generated by Russell Jurek
(ATNF) from 387 individual cubes. Its dimensions are 1721$\times$1721$\times$ 1025 with a data size of 12 GB.} as an example. With the expected increase in the output pixel resolution within the different main ASKAP HI surveys (e.g. WALLABY - expected output cube size of 6144 $\times$ 6144 $\times$ 16384), it will be useful to utilize such high resolution output facilities to provide the astronomers with a better level of detail for both qualitative and quantitative data visualization. 

On the other hand, such facilities may be useful to support collaborative visualization if a suitable data interaction mechanism is provided. We think the current framework can support an interactive OptIPortal display of the volume rendering output, but with a lower frame rate.

\section{Conclusion}
We demonstrate the ability to extend our GPU volume rendering framework to offer better visualization outcomes. Two main features were presented: integrating source finder output, and utilizing multi-panel displays. Both of these features demonstrate two main strong points in the framework design, namely the usage of GPU cluster as the main processing back-bone, and the separation between the rendering and the result display. We anticipate the ability to further enhance this for integrating more quantitative visualization tools and a better utilization of the GPU processing power.

\acknowledgements We thank Russell Jurek  (ATNF - CSIRO), and Virginia Kilborn (Swinburne) for providing sample data cubes.

\bibliographystyle{asp2010}
\bibliography{O11_3}

\begin{thebibliography}{}
\expandafter\ifx\csname natexlab\endcsname\relax\def\natexlab#1{#1}\fi
\expandafter\ifx\csname url\endcsname\relax
  \def\url#1{\texttt{#1}}\fi
\expandafter\ifx\csname urlprefix\endcsname\relax\def\urlprefix{URL }\fi
\providecommand{\eprint}[2][]{\url{#2}}

\bibitem[{Hassan et~al.(2011)Hassan, Fluke, \& Barnes}]{Hassan:2010a}
Hassan, A., Fluke, C., \& Barnes, D. 2011, New Astronomy, 16, 100

\bibitem[{Whiting(2008)}]{whiting:2008}
Whiting, M. 2008, Galaxies in the Local Volume, 343

\end{thebibliography}
\end{document}